\documentclass[%
reprint,
superscriptaddress,
showpacs,preprintnumbers,
 amsmath,amssymb,
 aps,
]{revtex4-1}

\usepackage{graphicx}
\usepackage{dcolumn}
\usepackage{bm}

\usepackage{epsfig}
\usepackage{hyperref}
\begin{document}

	\title{Dynamics of a magnetic skyrmionium in an anisotropy gradient}
	
	\author{Chengkun Song}%
	\affiliation{
		Key Laboratory for Magnetism and Magnetic Materials of the Ministry of Education, Lanzhou University, Lanzhou, 730000, People’s Republic of China
	}%
	\author{Chendong Jin}%
	\affiliation{
		Key Laboratory for Magnetism and Magnetic Materials of the Ministry of Education, Lanzhou University, Lanzhou, 730000, People’s Republic of China
	}%
	\author{Yunxu Ma}%
	\affiliation{
		Key Laboratory for Magnetism and Magnetic Materials of the Ministry of Education, Lanzhou University, Lanzhou, 730000, People’s Republic of China
	}%
	\author{Jinshuai Wang}%
	\affiliation{
		Key Laboratory for Magnetism and Magnetic Materials of the Ministry of Education, Lanzhou University, Lanzhou, 730000, People’s Republic of China
	}%
	\author{Haiyan Xia}%
	\affiliation{
		Key Laboratory for Magnetism and Magnetic Materials of the Ministry of Education, Lanzhou University, Lanzhou, 730000, People’s Republic of China
	}%

	\author{Jianbo Wang}%
	\affiliation{
		Key Laboratory for Magnetism and Magnetic Materials of the Ministry of Education, Lanzhou University, Lanzhou, 730000, People’s Republic of China
	}
	\affiliation{
		Key Laboratory for Special Function Materials and Structural Design of the of Ministry of Education, Lanzhou University, Lanzhou, 730000, People's Republic of China
	}%
	\author{Qingfang Liu}
	\email{liuqf@lzu.edu.edu}
	\affiliation{
		Key Laboratory for Magnetism and Magnetic Materials of the Ministry of Education, Lanzhou University, Lanzhou, 730000, People’s Republic of China
	}%
	
	
\date{\today}

\begin{abstract}
Magnetic skyrmionium is a novel magnetization configuration with zero skyrmion number, which is composed by two skyrmions with opposite skyrmion number. Here, we study the dynamics of skyrmionium under an anisotropy gradient. We find that the skyrmionium can be efficiently driven by an anisotropy gradient with moving straightly along the direction of gradient. The skyrmion Hall angle for skyrmionium is close to zero which is much smaller than that of skyrmion. while the speed is much larger. We also demonstrate that the skyrmionium motion depends on the damping cofficient, and the skyrmionium stabilization in the motion can be modulated by narrowing the width of the nanowire. Our work shows a efficient driven method for skyrmionium, which may be promising in the application of skyrmionium based racetrack memory. 
\end{abstract}


\maketitle 

Recently, magnetization configurations, such as droplets,~\cite{mohseni2013spin} domain walls (DW),~\cite{parkin2008magnetic,tatara2004theory} skyrmions~\cite{fert2013skyrmions,nagaosa2013topological} and skyrmioniums~\cite{zhang2016control,zhang2018real,li2018dynamics}, have attracted considerable attentions in the application of spintronic devices as information carriers. As topological protected magnetic structures, skyrmions are quite small in size and can be driven by spin transfer torque (STT) and spin Hall effect (SHE),~\cite{sampaio2013nucleation,jiang2015blowing,yu2012skyrmion,everschor2012rotating} the current density is significantly lower than that moving a magnetic DW. Since they have been theoretically predicted and experimently discovered in magnetic materials MnSi,~\cite{pappas2009chiral,neubauer2009topological} lots of researches are focused on the generation and manipulation of skyrmions.~\cite{lin2016edge,zhou2014reversible,liu2015skyrmion} They have been promised potential applications in next-generation memory and spintronic devices, such racetrack memory,~\cite{fert2017magnetic,sampaio2013nucleation,song2017skyrmion} spin transfer nano oscillators (STNOs),~\cite{zhang2015current,garcia2016skyrmion} logic devices,~\cite{zhang2015magnetic} neuromorphology devices~\cite{prychynenko2018magnetic,huang2017magnetic} and so on.

Skyrmion Hall effect is an important obstacle for the manipulation in confined magnetic geometries, which is induced by Magnus force acted on skyrmion.~\cite{jiang2017direct,litzius2017skyrmion} For skyrmions with topological number of $Q=\pm 1$, the deflection and corresponding skyrmion Hall angle are in opposite directions. Skyrmionium is a skyrmion-like magnetization configuration with a topological number $Q=0$.~\cite{zhang2018real,zhang2016control} It can be viewed as a composed structure nested by a skyrmion with $Q=1$ and a skyrmion with $Q=-1$. This intriguing characteristic makes skyrmionium as a promising information carrier for spintronic devices, where the dynamics are significantly different from those of skyrmions. The generation and stablization of skyrmionium have been theoretically proposed in the system with Dzyaloshinskii-Moriya interaction (DMI)~\cite{bogdanov1999stability,beg2015ground} or the gradient of curvature.~\cite{pylypovskyi2018chiral} It has been experimently verified in ferromagnetic thin film and ferromagnetic topological insulator heterostructure.~\cite{finazzi2013laser,zhang2018real} Recent investigations are focused on the skyrmionium manipulation, which can be driven by magnetic field,~\cite{komineas2015skyrmion} spin transfer torque~\cite{zhang2016control} and spin waves,~\cite{li2018dynamics} and the skyrmion Hall effect is limited for skyrmionium. However, a more energy efficient driving method is needed and the dynamics of skyrmioniums under anisotropy gradient still remain unexplored.

In this paper, we investigate the dynamics of a skrmionium under an anisotropy gradient as compared to skyrmions with opposite topological number $Q$. We also study the effect of damping constant. It is found that the skyrmionium can be efficiently driven along the direction of anisotropy gradient without deflection, and the moving velocity is obviously higher than that of skyrmions. Moreover, we show that the skyrmionium stabilization can be modulated by narrowing the nanowire width.

Micromagnetic simulations have been performed using Mumax3 code.\cite{vansteenkiste2014design} The time-dependent magnetization dynamics is given by the Landau-Lifshitz-Gilbert equation
\begin{equation}
\frac{\partial \mathbf{m}}{\partial t} = -\gamma \mathbf{m}\times \mathbf{H}_{\mathrm{eff}} + \alpha \mathbf{m} \times \frac{\partial \mathbf{m}}{\partial t},
\end{equation}
where $\mathbf{m} = \mathbf{M}/M_{\mathrm{s}}$ is the unit vector of the magnetization, $M_{\mathrm{s}}$ is the saturation magnetization. $\gamma$ and $\alpha$ are gyromagnetic ratio and the Gilbert damping, respectively. $\mathbf{H}_{\mathrm{eff}}=2A\nabla^2\mathbf{m}+2K_um_ze_z+\mathbf{H}_{\mathrm{DM}}$ is the effective field including the Heisenberg exchange field with exchange stiffness $A$, perpendicular magnetic anisotropy field with anisotropy coefficient $K_u$, and interfacial DMI field $\mathbf{H}_{\mathrm{DM}}$ characterized by DMI constant $D$. The ferromagnetic film size is 512 nm $\times$ 512 nm $\times$ 0.6 nm, the mesh size is 2 nm $\times$ 2 nm $\times$ 0.6 nm. The  model parameters are adopted as follows: $A=15 \times10^{-12} \ \mathrm{J/m}$, $K_u = 0.8\times10^6 \ \mathrm{J/m^3}$, $M_{\mathrm{s}}=580\times10^3 \ \mathrm{A/m}$. The Gilbert damping $\alpha$ varies from 0.01 to 0.3 and DMI strength $D=3.5 \ \mathrm{mJ/m^2}$.
\begin{figure}
	\begin{center}
		\epsfig{file=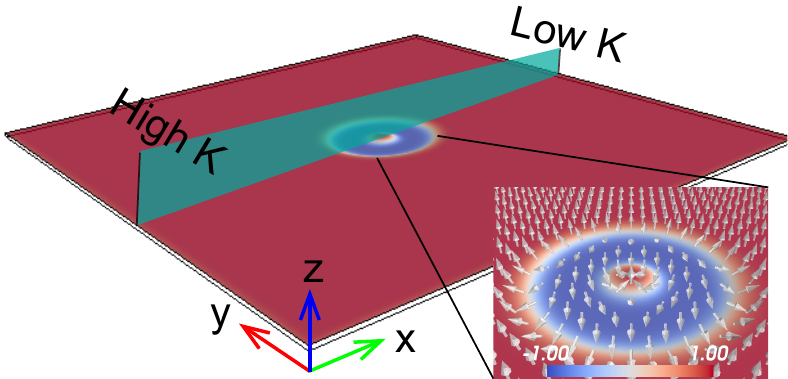,width=8 cm} \caption{
			Schematic of a VCMA gradient induced skyrmionium motion. A skyrmionium with positive polarity is located in a ferromagnetic thin film, and the inset shows the spatial distribution of the magnetization (the normalized $m_z$ represented by red color is positive and blue negative). The direction of anisotropy is along $x$-axis.}\label{fig1}
	\end{center}
\end{figure}

\begin{figure}
	\begin{center}
		\epsfig{file=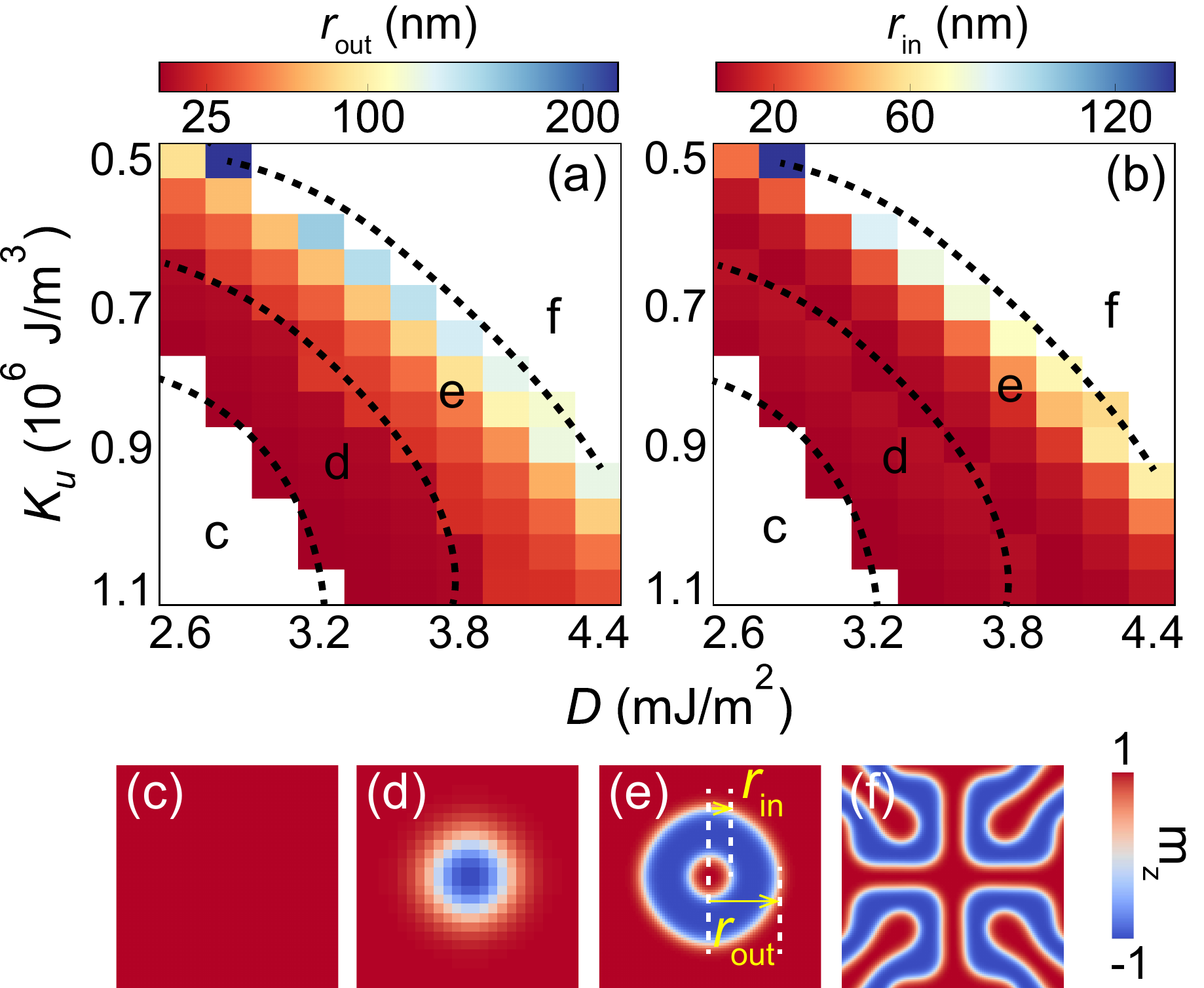,width=8 cm} \caption{Phase diagram of skyrmionium radius (a) $r_\mathrm{out}$ and (b) $r_\mathrm{in}$ as a function of $D$ and $K_u$. Four corresponding magnetization states are represented by (c) Single domain, (d) skyrmion, (e) skyrmionium and (f) labyrinth domain. $r_\mathrm{out}$ and $r_\mathrm{in}$ are defined in (e). The skyrmion radius for state (d) are simultaneously depicted in (a) and (b).}
		\label{fig2}
	\end{center}
\end{figure}

We consider that an electrode layer and a wedged insulating dielectric layer are fabricated on the top of the ferromagnetic layer, a constant linear anisotropy gradient is generated along the $x$-axis of system, as shown in Fig.~\ref{fig1}. The maximum (High $K$) and minimum (Low $K$) magnetic anisotropy are displayed at the edge of the film. Inset shows the magnetization configuration of a skrmionium located in the ferromagnetic film, where the red and blue colors represent positive and negative $z$-component of magnetization, respectively. The anisotropy distribution along $x$-axis is $K_{uv}=K_u + ({d K_u}/{dx}) x$, in which ${d K_u}/{dx}$ represents the anisotropy gradient, those values with High $K$ and Low $K$ are depicted in Table.~\ref{tab:table1}.
\begin{table}
	\caption{\label{tab:table1} Values of High and Low $K$ corresponding to the linear anisotropy gradient along $x$-axis. }
	\begin{ruledtabular}
		\begin{tabular}{lcr}
			High $K \ (\mathrm{J/m^3})$&Low $K \ (\mathrm{J/m^3})$&$\frac{d K_u}{dx} \ (\mathrm{J/m^4})$ \\
			\hline
			$0.800256\times10^6$ & $0.799744\times10^6$ & $-0.01\times10^{11}$\\
			$0.812800\times10^6$ & $0.787200\times10^6$ & $-0.5\times10^{11}$\\
			$\dots$ &  $\dots$& $\dots$ \\
			$0.825600\times10^6$  & $0.774400\times10^6$ & $-1.0\times10^{11}$\\
		\end{tabular}
	\end{ruledtabular}
\end{table}
Fig.~\ref{fig2} depicts the skyrmionium radius ($r_\mathrm{in}$ and $r_\mathrm{out}$) as a function of $K_u$ and $D$. The results show that the size increases with increasing $D$ and decreasing $K_U$. Four different magnetization states appear, which are single domain, skyrmion, skyrmionium and labyrinth domain. According to Table.~\ref{tab:table1} and Fig.~\ref{fig2}, the anisotropy variation along $x$-axis has a weak influence on skrmionium size change in the motion，thus we assume that the skyrmionium magnetization profile performs a translation without deformation.

First, we consider a skyrmionium placed in the ferromagnetic layer, which moves under anisotropy gradient. To describe the skyrmionium dynamics under anisotropy gradient, we use the model in the framework of the Thiele equation,~\cite{thiele1973steady} which treated the skyrmionium as a rigid particle,
\begin{equation}\label{eq2}
\mathbf{G}\times\mathbf{v}_s + \mathcal{D}\alpha \mathbf{v}_s=\mathbf{F},
\end{equation}
where $\mathbf{v}_s$ is the velocity of skyrmion, $\mathbf{G}=G\mathbf{e}_z = 4\pi Q \mathbf{e}_z$ is the gyromagnetic coupling vector. $Q=\frac{1}{4\pi}\int_m(\partial_x\mathbf{m}\times\partial_y\mathbf{m})dxdy$ is the skyrmion number. For skyrmionium with $Q=0$, the corresponding $\mathbf{G}=0$. $\mathcal{D}=
 \left(
 \begin{matrix}
   D_{ij} & 0 \\
   0 & D_{ij}
  \end{matrix}
  \right)
$ is the dissipative force tensor, where $D_{ij}=\int dxdy \partial_x\mathbf{m}\cdot \partial_x\mathbf{m}=\int dxdy \partial_y\mathbf{m}\cdot \partial_y\mathbf{m}$ with $i = j = x, y$. $\mathbf{F}$ is the driving force from the anisotropy gradient along $x$-axis described as
\begin{equation}
\mathbf{F}=\frac{\gamma}{\mu_0M_\mathrm{s}}\frac{d K_u}{dx}\int dxdy(1-m_z^2).
\end{equation}
The force due to boundary effect is not considered for that the skyrmionium is far from the edge of the sample. As the anisotropy gradient is only applied along $x$-axis, $\mathbf{F}=F\mathbf{e}_x$, we obtain
\begin{equation}
\begin{aligned}
Gv_y + \mathcal{D}\alpha v_x = & F, \\
-Gv_x + \mathcal{D}\alpha v_y = & 0 
\end{aligned}
\end{equation}
where $v_x$ and $v_y$ are the moving velocity in $x$ and $y$ directions. Thus
\begin{equation}\label{eq5}
\begin{aligned}
v_x &= \frac{\gamma}{\mu_0M_\mathrm{s}\alpha D_{ij}}\frac{d K_u}{dx}\int dxdy(1-m_z^2), \\
v_y &= 0
\end{aligned}
\end{equation}
which depicts that the skyrmionium motion is along the direction of anisotropy gradient $x$-axis, the velocity is proportinal to ${d K_u}/{dx}$ and inversely proportional to $\alpha$. 

While for skyrmion with $Q=\pm1$, $\mathbf{G}\ne 0$, the skyrmion velocity $v_x$ and $v_y$ are expressed as\cite{shen2018dynamics}
\begin{equation}
\begin{aligned}
 v_x &= \frac{d K_u}{dx}\frac{D_{ij}\alpha \gamma}{\mu_0M_\mathrm{s}((4\pi Q)^2+(\alpha D_{ij})^2)}\int dxdy(1-m_z^2) \\
v_y &= \frac{d K_u}{dx}\frac{4\pi Q \gamma}{\mu_0M_\mathrm{s}((4\pi Q)^2+(\alpha D_{ij})^2)}\int dxdy(1-m_z^2),
\end{aligned}
\end{equation}
thus the skyrmion velocity $v=\sqrt{v_x^2+v_y^2}=\frac{\gamma}{\mu_0M_\mathrm{s}\sqrt{(4\pi Q)^2+(\alpha D_{ij})^2}}\frac{d K_u}{dx}\int dxdy(1-m_z^2)$. Clearly, the skyrmionium velocity is larger than the skyrmion velocity with the same magnetic parameters. The moving directions for them are quite different. The skyrmion Hall angle is defined as 
\begin{equation}
\theta = \arctan(v_y/v_x),
\end{equation}
which is $\theta = 0$ for skyrmionium, and $\theta = \arctan((4\pi Q)/(D_{ij}\alpha))$ for skyrmion. Depending on the sign of $Q$ and the Magnus force acted on, skyrmions with opposite skyrmion number moves with opposite skyrmion Hall angle. As the skyrmionium is composed of two skyrmions with opposite skyrmion numbers ($Q=1$ and $Q=-1$). The total effective force acting on skyrmionium can be treated as the combination forces from two skyrmions, where the transverse force is zero.

\begin{figure}
	\begin{center}
		\epsfig{file=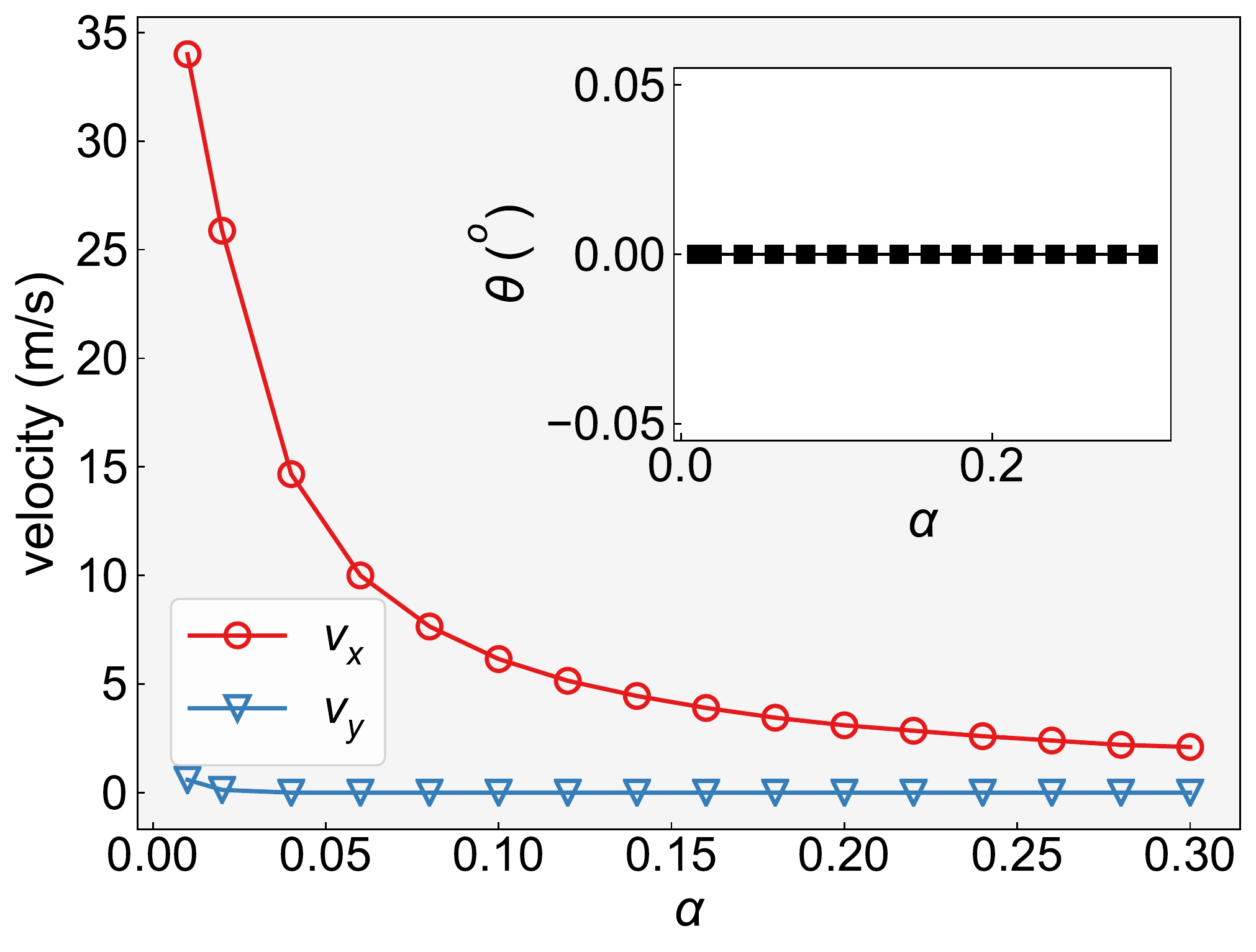,width=8 cm} \caption{
			Skyrmionium velocity $v_x$ (Red points) and $v_y$ (Blue points) as a function of $\alpha$, where $D=3.5 \ \mathrm{mJ/m^2}$ and $dK_u/dx = -0.5\times10^{11} \ \mathrm{J/m^4}$. The inset is skyrmion Hall angle $\theta$ for skyrmionium with varying $\alpha$.
			}\label{fig3}
	\end{center}
\end{figure}
Based on the analytical model, we carried out the simulation of the motion of skyrmionium. The simulated skyrmionium velocity $v_x$ and $v_y$ as a function of damping constant $\alpha$ are shown in Fig.~\ref{fig3}. A Inverse function relationship of $v_x$ related to $\alpha$ is observed as predicted, the corresponding $v_y$ is about zero. Inset in Fig.~\ref{fig3} shows the simulated results of $\theta$. It is worth noted that the skyrmionium size increases moving from high $K_u$ area to low $K_u$ area. While in our system, when we set the DMI strength $D=3.5 \ \mathrm{mJ/m^2}$ and $dK_u/dx = -0.5\times10^{11} \ \mathrm{J/m^4}$, the skyrmionium magnetization profile only exhibits small differeces between in the region of high $K$ and low $K$, and the change of skyrmionium size is negligible, as previously shown in Fig.~\ref{fig2} and Table.~\ref{tab:table1}.

\begin{figure}
	\begin{center}
		\epsfig{file=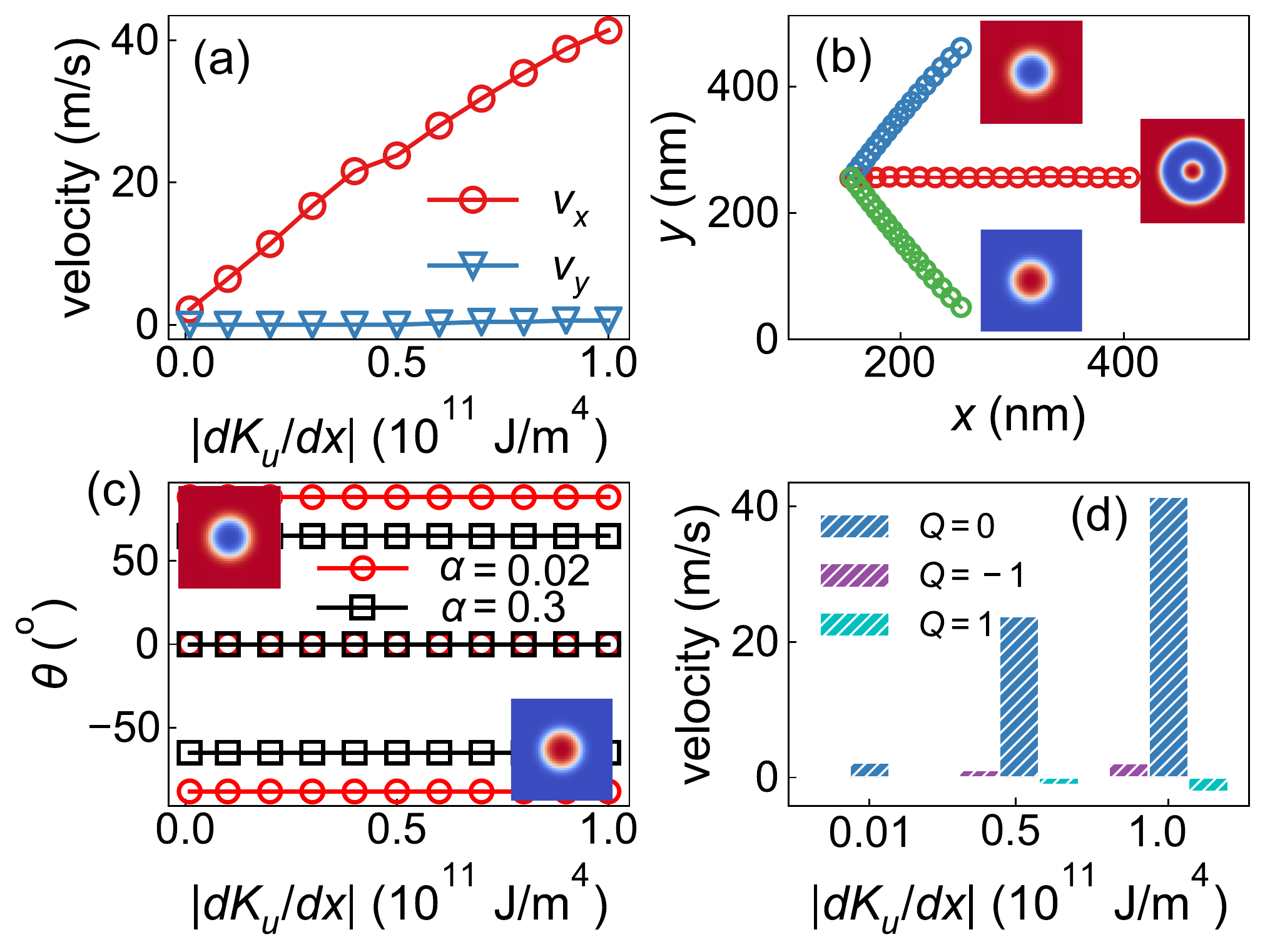,width=8 cm} \caption{
			(a) Velocities of Skyrmionium ($v_x$ and $v_y$) as a function of anisotropy gradient with $\alpha = 0.02$. (b) Trajectories of skyrmionium and skyrmion ($Q=\pm1$) driven by anisotropy gradient along $x$-direction with $\alpha = 0.3$. The dots denote the skyrmion and skyrmionium center. Insets are the snapshots of skyrmion with $Q=-1$, skyrmionium with $Q=0$ and skyrmion with $Q=1$ (from top to bottom). (c) Skyrmion Hall angle ($Q=-1, \ 0, \ 1$) as a function of anisotropy gradient with $\alpha=0.02$ and $\alpha = 0.3$. (d) Comparasion of velocities of skyrmion with $Q=\pm1$ and skyrmionium with $Q=0$ induced by anisotropy gradient.
			}\label{fig4}
	\end{center}
\end{figure}
In order to further investigate the skyrmionium dynamics driven by anisotropy gradient, we investigate its motion under different $dK_u/dx$, and compare with the dynamics of skyrmion with $Q=\pm1$, as shown in Fig.~\ref{fig4}. The results depicted in Fig.~\ref{fig4}(a) show that a larger anisotropy gradient will create a larger driving force which in turn lead a higher velocity for skyrmionium, the velocities of $v_x$ are linearly proportional to $|dK_u/d_x|$. However, $v_y$ is close to zero. These results are in consist with the analytical results in Eq.~\ref{eq5}. Fig.~\ref{fig4} (b) shows the trajectories of skyrmionium and skyrmions ($Q=\pm1$) at the damping parameter $\alpha=0.3$, where the anisotropy gradient is fixed as $dK_u/dx = -0.5\times10^{11} \ \mathrm{J/m^4}$. The skyrmionium straightly moves along the $+x$ direction without transverse motion. While for skyrmions, their motions exhibit a deflection towards the $+y$ and $-y$ directions for $Q=-1$ and $Q=1$, respectively. The transverse shift of skyrmion is referred as skyrmion Hall effect, which is similar to the skyrmion motion driven by current. A topological Magnus force will act on skyrmion when skyrmion is moving, which is perpendicular to the velocity of skyrmion. The directions of Magnus force depends on sign of the skyrmion number $Q$, which are opposite for $Q=-1$ and $Q=1$. As the skyrmionium can be viewed as a hybrid of skyrmion with opposite $Q$, the Magnus force acted on skyrmionium is cancelled out, thus the skyrmionium can moves in a straight trajectory. 

Figure.~\ref{fig4} (c) shows the skyrmion Hall angle $\theta$ as a function of anisotropy gradient for $\alpha=0.02$ and $\alpha=0.3$. The skyrmionium moves straightly driven by anisotropy gradient and $\theta = 0^o$. Moreover, the skyrmion Hall angle for skyrmionium is independent of damping constant. However, skyrmion with $Q=-1$ and $Q=1$ exhibit a large skyrmion Hall angle $\theta$. For skyrmion with $Q=-1$ (upper inset), $\theta=88.25^o$ with $\alpha=0.02$, and it decreases to $65^o$ with $\alpha=0.3$. While for skyrmion with $Q=1$ (lower inset), $\theta$ is $-88.25^o$ and $-65^o$ for $\alpha = 0.02$ and $0.3$, respectively. By comparing the simulated results of skyrmionium and skyrmion velocities at certain values of $|dK_u/dx|$ for $\alpha=0.02$ (Fig.~\ref{fig4} (d)), where $|dK_u/dx| = 0.01, \ 0.5, \ 1.0 \times10^{11} \ \mathrm{J/m^4}$, it is found that the skyrmionium velocity is much larger than that of skyrmion. It is worth mentioning that only $v_x$ of skyrmionium and $v_y$ of skyrmions are depicted to demonstrate the moving directions of skyrmions in Fig.~\ref{fig4}, for that $v_y=0$ for skyrmionium and $v_x$ for skyrmions are quite small.    

\begin{figure}
	\begin{center}
		\epsfig{file=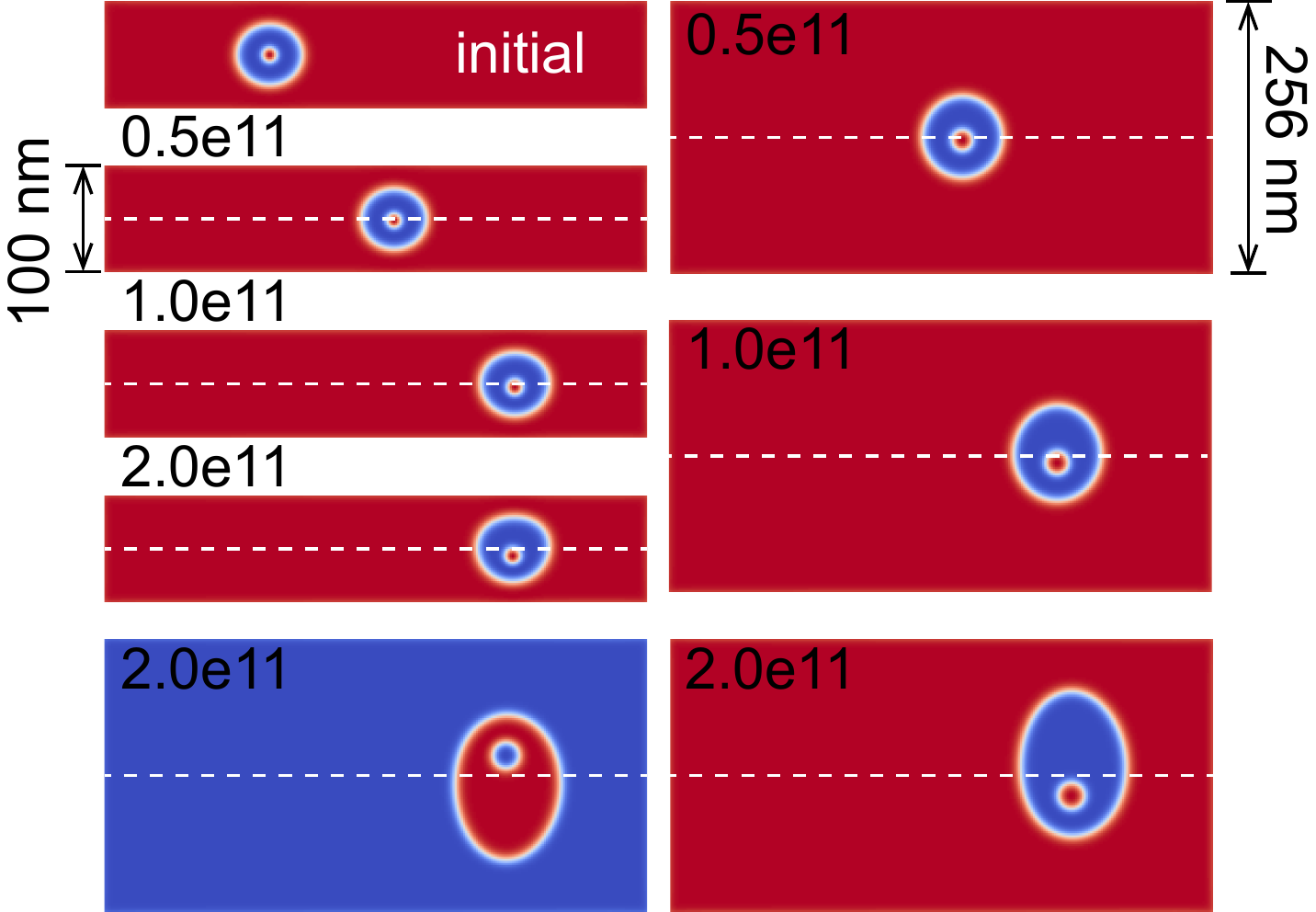,width=8 cm} \caption{
			Comparasion of skyrmionium motion driven by anisotropy gradient in nanowires with different width, $d=100$ nm (left column) and $d = 256$ nm (right column), the anisotropy gradient is $|dK_u/dx| = 0.5, \ 1.0, \ 2.0 \times10^{11} \ \mathrm{J/m^4}$, respectively. The white dashed lines indicate the center position along $y$ direction.
			}\label{fig5}
	\end{center}
\end{figure}
A larger anisotropy gradient is necessary to increase skyrmionium velocity, while it induces a significant change of skyrmionium size, which limits the application in racetrack memory. The skyrmionium deformation with moving in nanowire under a larger anisotropy at $\alpha=0.02$ is indicated in Fig.~\ref{fig5}. When the skyrmionium moves in a nanowire with width $d = 256$ nm, its magnetization profile keeps unchanged in a small anisotropy gradient with $|dK_u/dx| = 0.5 \times10^{11} \ \mathrm{J/m^4}$. When the anisotropy gradient increases to $1 \times10^{11} \ \mathrm{J/m^4}$, the skyrmionium magnetization profile only exhibits a slight change with a small transverse shift for inner part, and the outer part elongates along $y$ direction. While for $|dK_u/dx| = 2 \times10^{11} \ \mathrm{J/m^4}$, the transverse shift of inner part and the elongation of outer part for skyrmionium is more significant. Two nested skyrmions experience the Magnus force when the skyrmionium is moving, which are in opposite directions ($+y$ and $-y$). In a large anisotropy gradient, the Magnus force acted on two skyrmions increases with increasing speed. Thus, the skyrmionium stabilization is perturbed with the inner and outer skyrmion moving toward $-y$ and $y$ directions, respectively. The competetion between these two effect induces the skyrmion deformation in the large anisotropy. As depicted in the left corner of Fig.~\ref{fig5}, a skyrmionium with negative polarity moves with a deformation that exhibits a slight change with a small transverse shift along $y$ direction for inner part, which is opposite to that shown in the right corner of Fig.~\ref{fig5}. 

With decreasing the width of nanowire to 100 nm, the skyrmionium size for initial state reduces compared to the skyrmionium in the nanowire with $d=256$ nm. In different anisotropy gradients, the skyrmionium magnetization profiles keep stable, and a small deformation appear in a larger anisotropy gradient $|dK_u/dx| = 2 \times10^{11} \ \mathrm{J/m^4}$. The Magnus force acted on inner and outer part of skyrmionium is compensated by the edge of nanowire, and the deformation is limited comparing to the case of $d=256$ nm. It is found that the boundary confinment of the nanowire plays an important role to stabilize skyrmionium. Many methods have been proposed to confining the skyrmion in the racetrack and increasing moving speed, such as modifying magnetic parameters (magnetic anisotropy, DMI),~\cite{mulkers2017effects,upadhyaya2015electric} using curbed nanowire.~\cite{purnama2015guided} By modification the boundary effect of skyrmionium based racetrack, the stabilizaiton and moving speed can be enhanced.

In summary, we have studied the skyrmionium motion in an anisotropy gradient. We find that the skyrmionium moves straightly along the direction of anisotropy gradient, and the velocity increases linearly with $dK_u/dx$ which is much larger than that of magnetic skyrmion. Moreover, we find that a larger anisotropy gradient induces deformation of skyrmionium. By narrowing the nanowire and enhancing the boundary confinment, the skyrmionium deformation can be limited with a large motion speed. Our results may be useful in the application of energy efficient skyrmionium based spintronic devices.

This work is supported by National Natural Science Fund of China (Grants No. 11574121 and No. 51771086)

\bibliography{reference}

\end{document}